\documentclass[prd,twocolumn,showpacs,amsmath,amssymb]{revtex4}
\usepackage{graphicx}
\usepackage{amssymb}
\begin{document}

\title{Application of higher order holonomy corrections to perturbation theory of cosmology}
\author{Yu Li}
\email{leeyu@mail.bnu.edu.cn}
  \affiliation{Department of Physics, Beijing Normal University, Beijing 100875, China}
\author{Jian-Yang Zhu}
 \email{zhujy@bnu.edu.cn}
  \affiliation{Department of Physics, Beijing Normal University, Beijing 100875, China}

\begin{abstract}
Applying the higher order holonomy corrections to the perturbation
theory of cosmology, the lattice power law of Loop Quantum
Cosmology, $\tilde{\mu}\propto p^{\beta}$, is analysed and the range
of $\beta$ is decided to be [-1,0] which is different from the
conventional range $-0.1319>\beta\geq-5/2$ \cite{lqct}. At the same
time, we find that there is a anomaly free condition in this theory,
and we obtain this condition in the vector and tensor mode. We also
find that the nonzero mass of gravitational wave essentially results
from the quantum nature of Riemannian geometry of loop quantum
gravity.

\end{abstract}

\maketitle

\section{Introduction}

The spacetime metric of Big Bang cosmology is homogeneous and
isotropic Friedmann-Robertson-Walker (FRW) metric. However, this
model is just an approximation of ``zero order" universe \cite{mc}.
If we only focus on the FRW metric, we will ignore many of
interesting things in the universe such as galaxy clusters,
galaxies, stars, etc. So it is necessary to introduce the
inhomogeneous and anisotropy perturbation to describe these things
\cite{w}. On the other hand, the effects of quantum gravity should
be significant in the very early universe. Therefore, it is
interesting to study possible quantum gravity effects in
cosmological perturbation theory.

At present, the problem of finding the quantum theory of the
gravitational field is still open. One of the most active of the
current approaches is loop quantum gravity. Loop quantum gravity
(LQG) \cite{lqg1,lqg2,lqg3} is a mathematically well-defined,
non-perturbative and background independent quantization of general
relativity. Its cosmological version, the loop quantum cosmology
(LQC) \cite{B} have achieved many successes. A major success of LQC
is the resolution of the Big Bang singularity \cite{bb,nbb1,nbb2};
this result depends crucially on the discreteness of the spacetime
geometry. With such a result, the big-bang singularity will be
avoided through a big-bounce mechanism in the high energy region. In
addition, LQC can also setup suitable initial conditions for
successful inflation \cite{if1,if2} as well as possibly leaving an
imprint in the cosmic microwave background \cite{if2}.

In LQG, spacetime is quantized.  The geometric operators, such as
the area operator and the volume operator, have discrete
eigenvalues. So there is the smallest area gap $\Delta $
\cite{sa1,sa2}. In LQC, the coordinate size of a loop is
$\tilde{\mu}^2$. $\tilde{\mu}$ is the function of $p=a^2$ (where $a$
is the scale factor of the universe.), i.e.
$\tilde{\mu}=\tilde{\mu}(p)$. In the early literature
\cite{nbb1,mslqc}, the work always base on the simplest choice of
$\tilde{\mu}(p)=\mu_0=const$. However, this form can lead to some
unusual features. As pointed out in \cite {nbb2} that the choice of
$\tilde{\mu}(p)=const$ can lead to the Big Bounce occurs at
classical matter density like water, so he suggest to select the
function as $\tilde{\mu}(p)\propto p^{-1/2}$. From this time forth,
in most of current works \cite{jcap1}, $\tilde{\mu}(p)\propto
p^{-1/2}$ has been applied. And it was shown that  the choice
$\tilde{\mu}(p)\propto p^{-1/2}$ is physically and mathematically
consistent \cite{nbb2}. Up to now, however, there is still no theory
to decide the function of $\tilde{\mu}(p)$. As research continues,
there may be some other form of $\tilde{\mu}(p)$ can give the better
physics. Therefore, to find out the form of this function has
theoretical significance.

An ansatz for the form of this function can be taken as $\tilde{\mu}(p)\propto p^{\beta}$.
In \cite{lqct}, the range of $%
\beta$ has been decided to be $-0.1319>\beta\geq-5/2$. However, it
is just the conclusion of the first order holonomy corrections. If
we want a more accurate determination of
 the range of $\beta$, we must consider the higher
order corrections.

Even in the case of homogeneous and isotropic models, the quantum equation of state is very difficult to analyze.
Fortunately, there is a powerful tool, i.e.
effective theory, which allows us to include loop quantum effects
by correction terms in equations of the classical type \cite{blqc}.
There are two types of quantum corrections that are expected from
the Hamiltonian of LQG. One correction arises for
inverse powers of the densitized triad, which when quantized becomes
an operator with zero in the discrete part of its spectrum thus
lacking a direct inverse. The other comes from the fact that a loop
quantization is based on holonomies, i.e. exponentials of the
connection rather than direct connection components \cite{thie}.

In LQC, there is no well-defined quantum operator corresponding to
$c=\gamma k $. So we should find a well-defined operator to replace
it. The conventional way is replacing the $c$ by
$\sin{\tilde{\mu}c}/\tilde{\mu}$.

The application of inverse triad corrections and conventional
holonomy corrections on the scalar mode of perturbation can be viewed in \cite{lqcs},
the vector mode in \cite {lqcv} and the tensor mode in \cite{lqct}.

In this paper, we focus on the higher holonomy corrections rather
than the conventional correction. We apply these higher corrections
to the vector and the tensor mode and see whether the mass of
gravitational wave is the nature of discrete geometry. We will also
analyses the range of $\beta$ with high order holonomy corrections.

This paper is organized as follows. At first, the perturbed
variables are introduced in Sec. \ref{s2}. And then in Sec.
\ref{s4}, we apply the high order holonomy corrections to obtain the
effective Hamiltonian constraint. Detailed analysis of the range of
$\beta $ will be given in Sec. \ref {s5}. Section \ref{s6} is our
discussion.

\section{Background and perturbed constraint}

\label{s2} In Ashtekar's formalism of general relativity
\cite{af,af2}, the
spatial metric as a canonical field is replaced by the densitized triad $%
E_i^a$, defined as
\begin{equation}
E_i^a:=\left| \det \left( e_b^j\right) \right| e_i^a,  \label{1}
\end{equation}
and the spin connection $\Gamma _a^i$  which is
\begin{equation}
\Gamma _a^i=-\epsilon ^{ijk}e_j^b\left( \partial _{[a}e_{b]}^k+\frac 12%
e_k^ce_a^l\partial _{[c}e_{b]}^l\right) .  \label{2}
\end{equation}

The canonical variables are densitized triad $E^a_i$ and Ashtekar
connection
$A_a^i=\Gamma^i_a+\gamma K^i_a$, where $K^i_a$ is extrinsic curvature and $%
\gamma$ is Barbero-Immirzi parameter.

The canonical variables reduced to spatially homogeneous and
isotropic cosmology are
\begin{equation}
E_i^a=p\delta _i^a,\ \ \ K_a^i=k\delta _a^i,\ \ \ \Gamma _a^i=0.
\end{equation}
They are background variables, and the perturbation will be added
based on these variables.

In the perturbation theory, we denote the background variables by a
bar:
\begin{equation}
\bar{E}_i^a=\bar{p}\delta _i^a,\ \ \ \bar{\Gamma}_a^i=0,\ \ \ \bar{K}_a^i=%
\bar{k}\delta _a^i,\ \ \ \bar{N}=\sqrt{\bar{p}};\ \ \ \bar{N}^a=0,
\label{3}
\end{equation}
where $\bar{p}=a^2$, and the spatial metric is
$\bar{q}_{ab}=a^2\delta _{ab}$. We use conformal time in this paper,
so we set $\bar{N}=a$.

The canonical variables are the perturbation densitized triad
$E_i^a$ and and Ashtekar connection $A_a^i$, which are
\begin{equation}
E_i^a=\bar{p}\delta _i^a+\delta E_i^a,\ \ A_a^i=\Gamma _a^i+\gamma
K_a^i=\gamma \bar{k}\delta _a^i+\left( \delta \Gamma _a^i+\gamma
\delta K_a^i\right) ,  \label{4}
\end{equation}
where $\delta E_i^a$ and $\delta K_a^i$ are small perturbation
around homogeneous variables.

As described in \cite{lqcs,lqcv,lqct}, the symplectic structure
splits into two parts: one for the background variables and the
other for perturbations, i.e.,
\begin{equation}
\left\{ \bar{k},\bar{p}\right\} =\frac{8\pi G}{3V_0},\label{5}
\end{equation}
and
\begin{eqnarray}
\left\{ \delta K_a^i\left( x\right) ,\delta E_j^b\left( y\right)
\right\} &=&8\pi G\delta ^3\left( x,y\right) \delta _a^b\delta
_j^i.\label{55}
\end{eqnarray}
Here, $G$ is the gravitational constant and $V_0$ is a fiducial
volume.

In vector mode, the gravity part of perturbed Hamiltonian constraint
(up to quadratic terms) is \cite{lqcv}
\begin{eqnarray}
H_G\left[ N\right]  &=&\frac 1{16\pi G}\int_\Sigma d^3x\bar{N}\left[ \bar{k}%
^2\left( -6\sqrt{\bar{p}}-\frac{\delta E_j^c\delta E_k^d\delta
_c^k\delta
_d^j}{2\bar{p}^{3/2}}\right) \right.   \nonumber  \label{6} \\
&&\left. +\sqrt{\bar{p}}\left( \delta K_c^j\delta K_d^k\delta
_k^c\delta _j^d\right) -\frac{2\bar{k}}{\sqrt{\bar{p}}}\left( \delta
E_j^c\delta K_c^j\right) \right] .
\end{eqnarray}
On the other hand, when we introduce the inhomogeneous perturbation,
the diffeomorphism constraint does not vanish any more. So the
gravitational part of diffeomorphism constraint is changed into
\cite{lqcv}
\begin{equation}
D_G\left[ N^a\right] =\frac 1{8\pi G}\int_\Sigma d^3x\delta N^c\left[ -\bar{p%
}\left( \partial _k\delta K_c^k\right) -\bar{k}\delta _c^k\left(
\partial _d\delta E_k^d\right) \right] .  \label{7}
\end{equation}
Using Eqs.(\ref{5}) and (\ref{55}), we can testify the following
relation easily
\begin{equation}
\left\{ H_G,D_G\right\} =0.  \label{8}
\end{equation}

Similarly, in tensor mode, the gravity part of the perturbed Hamiltonian
constraint (up to quadratic terms) is \cite{lqct}
\begin{eqnarray}
H_G[N] &=&\frac 1{16\pi G}\int_\Sigma d^3x\bar{N}\left[ \bar{k}^2\left( -6%
\sqrt{\bar{p}}-\frac{\delta E_j^c\delta E_k^d\delta _c^k\delta _d^j}{2\bar{p}%
^{3/2}}\right) \right.   \nonumber \\
&&+\sqrt{\bar{p}}\left( \delta K_c^j\delta K_d^k\delta _k^c\delta
_j^d\right) -\frac{2\bar{k}}{\sqrt{\bar{p}}}(\delta E_j^c\delta
K_c^j)
\nonumber \\
&&\left. +\frac 1{\bar{p}^{3/2}}\left( \delta _{cd}\delta
^{jk}\delta ^{ef}\partial _eE_j^c\partial _fE_k^d\right) \right] , \label{9}
\end{eqnarray}
where $\delta E_i^a=-\frac 12\bar{p}h_i^a$, here $h_a^i:=\delta ^{ib}h_{ab}$%
, and $h_{ab}$ is the symmetric metric perturbation field. It is
transverse and traceless, i.e. it satisfies $\partial ^ah_{ab}=0$
and $\delta ^{ab}h_{ab}=0$ \cite{w}.

\section{Vector and tensor mode with higher order holonomy corrections\label{s4}}

\subsection{Higher order holonomy corrections}
Instead of the conventional way of introducing the holonomy
corrections, in this article, we focus on the higher order holonomy corrections
\cite{hhc}.

At first, let's consider the Taylor
series
\begin{equation}
\sin^{-1}x=\sum_{l=0}^{\infty}\frac{\left( 2l\right) !}{%
2^{2l}\left( l!\right) ^2\left( 2l+1\right) }x^{2l+1}\label{40}
\end{equation}
for $-1\leq x\leq 1$ and setting $x=\sin\left(\tilde{\mu}\gamma \bar{k}\right)$, we have
\begin{equation}
\gamma \bar{k}=\frac 1{\tilde{\mu}}\sum_{l=0}^{\infty}\frac{\left( 2l\right) !}{%
2^{2l}\left( l!\right) ^2\left( 2l+1\right) }\left[ \sin
\left(\tilde{\mu}\gamma \bar{k}\right)\right] ^{2l+1},\label{41}
\end{equation}
This inspires us to define a $n$th order holonomized connection variable as
\begin{equation}
c_h^{(n)}:=\frac 1{\tilde{\mu}}\sum_{l=0}^n\frac{\left( 2l\right) !}{%
2^{2l}\left( l!\right) ^2\left( 2l+1\right) }\left[ \sin
\left(\tilde{\mu}\gamma \bar{k}\right)\right] ^{2l+1},  \label{11}
\end{equation}
which can be made arbitrarily close to $\gamma \bar{k}$ as
$n\rightarrow \infty$. We can see that $c_h^{(n)}$ is a function of
the holonomy $\sin \left(\tilde{\mu}\gamma \bar{k}\right)$ and the
discreteness variable $\tilde{\mu}$. Therefore, we can replace
$\gamma \bar{k}$ by $c_h^{(n)}$ to implement the underlying
structure of LQC. When $n=0$, $c_h^{(0)}=\sin\left(\tilde{\mu}\gamma
\bar{k}\right)/\tilde{\mu}$ is the same with the conventional
holonomy corrections.

There is an ambiguity in this replacement. If we set $x=\sin\left(m\tilde{\mu}\gamma \bar{k}\right)$ in Eq.(\ref{40}),
where $m$ is an arbitrary constant, Eq.(\ref{41}) changes to
\begin{equation}
 m\gamma\bar{k}=\frac 1{\tilde{\mu}}\sum_{l=0}^{\infty}\frac{\left( 2l\right) !}{%
2^{2l}\left( l!\right) ^2\left( 2l+1\right) }\left[ \sin
\left(m\tilde{\mu}\gamma \bar{k}\right)\right] ^{2l+1},
\end{equation}
and we have
\begin{equation}
\gamma \bar{k}=\frac 1{m\tilde{\mu}}\sum_{l=0}^{\infty}\frac{\left( 2l\right) !}{%
2^{2l}\left( l!\right) ^2\left( 2l+1\right) }\left[ \sin
\left(m\tilde{\mu}\gamma \bar{k}\right)\right] ^{2l+1}.
\end{equation}
So we can define a more general $n$th order holonomized connection
variable $c_{mh}^{(n)}$:
\begin{equation}
c_{mh}^{(n)}:=\frac 1{m\tilde{\mu}}\sum_{l=0}^n\frac{\left( 2l\right) !}{%
2^{2l}\left( l!\right) ^2\left( 2l+1\right) }\left[\sin\left(
m\tilde{\mu}\gamma \bar{k}\right)\right] ^{2l+1},  \label{12}
\end{equation}
where $m$ is an ambiguity parameter.

The Poisson brackets between the canonical variables and the
$c_h^{(n)}$ are
\begin{equation}
\left\{ \bar{p},\frac{c_h^{(n)}}\gamma \right\} =-\frac{8\pi
G}{3V_0}\cos
\left( \tilde{\mu}\gamma \bar{k}\right) \mathfrak{G}_n\left( \tilde{\mu}%
\gamma \bar{k}\right) ,
\end{equation}
and
\begin{equation}
\left\{ \bar{k},\frac{c_h^{(n)}}\gamma \right\} =\frac{8\pi G}{3\tilde{\mu}%
V_0}\frac{\partial \tilde{\mu}}{\partial \bar{p}}\left[ \cos (\tilde{\mu}%
\gamma \bar{k})\mathfrak{G}_n\left( \tilde{\mu}\gamma \bar{k}\right) \bar{k}-%
\frac{c_h^{(n)}}\gamma \right] ,
\end{equation}
where
\begin{equation}
\mathfrak{G}_n(\tilde{\mu}\gamma \bar{k})=\sum_{l=0}^n\frac{\left(
2l\right)
!}{2^{2l}\left( l!\right) ^2}\left[ \sin \left( \tilde{\mu}\gamma \bar{k}%
\right) \right] ^{2l}.  \label{14}
\end{equation}

From \cite{hhc} we can see that the role of higher order holonomy
corrections is like a filter, which excludes the impact of human
factors on the theory and leaves a pure quantum effect.

\subsection{Vector mode}

In classical perturbation theory of cosmology, the gauge invariant
variables of the vector mode  will decay quickly. Therefore, there
is a little role of the vector mode perturbation for a universe
\cite{w}. However, once we introduce the quantum correction, we must
consider whether the perturbation theory is anomaly free
\cite{lqcv}. The requirement of anomaly free can reduce some
ambiguities of LQC. Inserting the higher holonomy corrections in
Eq.(\ref{6}), we can obtain the effective gravity part of the
perturbed Hamiltonian constraint
\begin{eqnarray}
H_G^Q[N] &=&\frac 1{16\pi G}\int_\Sigma d^3x\bar{N}\left\{ \left( \frac{%
c_h^{(n)}}\gamma \right) ^2\right.   \nonumber \\
&&\times \left[ -6\sqrt{\bar{p}}-\frac 1{2\bar{p}^{3/2}}\left(
\delta
E_j^c\delta E_k^d\delta _c^k\delta _d^j\right) \right]   \nonumber \\
&&\left. +\sqrt{\bar{p}}\left( \delta K_c^j\delta K_d^k\delta
_k^c\delta _j^d\right) -\frac
2{\sqrt{\bar{p}}}\frac{c_{mh}^{(n)}}\gamma \delta E_j^c\delta
K_c^j\right\} .  \nonumber \\
\end{eqnarray}

General speaking, we should replace all the $\gamma \bar{k}$ by $%
c_{mh}^{(n)}$. But in order to get a homogeneous limit which
agreement with what has been used in isotropic models, we set the
parameter $m$ in the first term to equal one \cite{lqct}. The
parameter $m$ in the last term should lead to an anomaly-free
constraint algebra, so we do not fix it at first. In the following discussion,
we will determine  the
right value of $m$ in the last term by requiring an anomaly-free
constraint algebra in the presence of quantum corrections.

In homogeneous and isotropic model, there is no diffeomorphism
constraint. So the algebra of constraints is closed. When we
consider the inhomogeneous perturbation, the diffeomorphism
constraint will turn up. From Eq.(\ref {8}) we can see that, in
classical theory, the algebra of constraints is still closed. So
when we write down the constraints with the quantum corrections, we
need to ensure that the constraints are still closed. In other
words, the anomaly terms, which cannot be expressed by the linear
combination of the Hamiltonian constraint and the diffeomorphism
constraint, should be vanished. On the other hand, the
diffeomorphism constraint does not receive quantum corrections in
the full theory \cite{af1}, so Eq.(\ref{7}) does not change.

The Poisson bracket between two constraints is
\begin{eqnarray}
&&\{H_G^Q,D_G\}  \nonumber \\
&=&\frac{\bar{N}}{\sqrt{\bar{p}}}\left[ \bar{k}+\frac{%
c_{mh}^{(n)}}\gamma -2\frac{c_h^{(n)}}\gamma \cos \left(
\tilde{\mu}\gamma \bar{k}\right) \mathfrak{G}_n\left(
\tilde{\mu}\gamma \bar{k}\right)
\right] D_G  \nonumber  \label{16} \\
&+&\frac 1{8\pi G}\int_\Sigma d^3x\bar{p}(\partial _c\delta N^j){\cal A}%
_j^{(n)c},
\end{eqnarray}
where the anomaly part is
\begin{eqnarray}
{\cal A}_j^{(n)c} &=&\frac{\bar{N}}{\sqrt{\bar{p}}}\left\{ \bar{p}\frac %
\partial {\partial \bar{p}}\left( \frac{c_h^{(n)}}\gamma \right) ^2+\left(
\frac{c_h^{(n)}}\gamma \right) ^2-\bar{k}^2\right.   \nonumber  \label{17} \\
&&\left. +2\left[ \frac{c_h^{(n)}}\gamma \cos (\tilde{\mu}\gamma \bar{k})%
\mathfrak{G}_n(\tilde{\mu}\gamma \bar{k})-\frac{c_{mh}^{(n)}}\gamma
\right] \bar{k}\right\} \left( \frac{\delta E_j^c}{\bar{p}}\right) .  \nonumber \\
\end{eqnarray}
To get this, we need the Poisson bracket
\begin{eqnarray}
\left\{ \delta K_c^j(x),\partial _d\delta E_k^d(y)\right\}  &=&8\pi
G\delta
_k^j\delta _c^d\partial _d\delta (x,y),  \label{18} \\
\left\{ \delta E_c^j(x),\partial _d\delta K_k^d(y)\right\}  &=&-8\pi
G\delta _k^j\delta _c^d\partial _d\delta (x,y).
\end{eqnarray}
To cancel the anomaly part, it must be requested that ${\cal A}_j^{(n)c}=0$, i.e.
\begin{eqnarray}
\frac{c_{mh}^{(n)}}{\gamma \bar{k}} &=&\left( \beta +1\right) \frac{c_h^{(n)}%
}{\gamma \bar{k}}\cos \left( \tilde{\mu}\gamma \bar{k}\right) \mathfrak{G}%
_n\left( \tilde{\mu}\gamma \bar{k}\right)   \nonumber \\
&&+\frac{1-2\beta }2\left( \frac{c_h^{(n)}}{\gamma \bar{k}}\right)
^2-\frac 12.  \label{19}
\end{eqnarray}

From \cite{hhc} we know that the big bounce occur when $\tilde{\mu}c=\frac \pi 2$%
, so the maximum of $c_{mh}^{(n)}\tilde{\mu}$ is $\frac \pi 2$.
According Eq.(\ref{19}), we have
\begin{widetext}
\begin{equation}
\left( \beta +1\right) c_h^{(n)}\tilde{\mu}\cos \left(
\tilde{\mu}\gamma \bar{k}\right) \mathfrak{G}_n\left(
\tilde{\mu}\gamma \bar{k}\right) +\left[
\frac{1-2\beta }2\left( \frac{c_h^{(n)}}{\gamma \bar{k}}\right) ^2-\frac 12%
\right] \tilde{\mu}\gamma \bar{k}\leq \frac \pi 2.  \label{32}
\end{equation}
\end{widetext}

Eq.(\ref{32}) can be seen as a limit to the evolution of
$c_{mh}^{(n)}$, and we can restrict the range of $\beta$ through
this limit.

\subsection{Tensor mode}

In classical perturbation theory of cosmology, there is only one
equation in tensor mode, i.e. the gravitational waves equation. From
this equation, we know that the gravitational waves are massless.
However, when quantum corrections are taken into account,
a mass term will be appeared in this equation \cite{lqct}. It is
only the conclusion calculated in the first order correction. We
extend this method to the higher holonomy corrections, and
take limit of $n\rightarrow \infty$. In this way, we can find
that the mass of gravitational waves is the intimately results from
the quantum nature of Riemannian geometry of LQG.

Inserting the higher holonomy corrections to Eq.(\ref{9}), the
effective gravity part of perturbed Hamiltonian constraint can be
expressed as
\begin{eqnarray}
H_G^Q[N] &=&\frac 1{16\pi G}\int_\Sigma d^3x\bar{N}\left\{ \left( \frac{%
c_h^{(n)}}\gamma \right) ^2\left[ -6\sqrt{\bar{p}}\right. \right.
\nonumber
\\
&&\left. -\frac 1{2\bar{p}^{3/2}}\left( \delta E_j^c\delta
E_k^d\delta _c^k\delta _d^j\right) \right] +\sqrt{\bar{p}}\left(
\delta K_c^j\delta
K_d^k\delta _k^c\delta _j^d\right)   \nonumber \\
&&\left. -\frac 2{\sqrt{\bar{p}}}\frac{c_{mh}^{(n)}}\gamma \delta
E_j^c\delta K_c^j+\frac{\delta _{cd}\delta ^{jk}\delta ^{ef}\partial
_eE_j^c\partial _fE_k^d}{\bar{p}^{3/2}}\right\} . \nonumber \\
\end{eqnarray}
From this Hamiltonian, one can obtain the time derivative of the
background variables
\begin{eqnarray}
\bar{p} &=&2\bar{p}\frac{c_h^{(n)}}\gamma \cos \left( \tilde{\mu}\gamma \bar{%
k}\right) \mathfrak{G}_n\left( \tilde{\mu}\gamma \bar{k}\right)   \label{21}
\\
\bar{k} &=&-\frac 12\left( \frac{c_h^{(n)}}\gamma \right) ^2-2\frac{c_h^{(n)}%
}\gamma \frac{\bar{p}}{\tilde{\mu}}\frac{\partial \tilde{\mu}}{\partial \bar{%
p}}  \nonumber \\
&&\times \left[ \cos \left( \tilde{\mu}\gamma \bar{k}\right) \mathfrak{G}%
_n\left( \tilde{\mu}\gamma \bar{k}\right) \bar{k}-\frac{c_h^{(n)}}\gamma
\right] ,
\end{eqnarray}
and the time derivative of the perturbed variable
$\delta E_i^a$
\begin{equation}
\dot{\delta E_i^a}=\left\{ \delta E_i^a,H_G^Q\right\}
=-\bar{p}\delta _k^a\delta _i^d\delta K_d^k-\frac
12\bar{p}\frac{c_{mh}^{(n)}}\gamma h_i^a. \label{22}
\end{equation}
On the other hand, one can also obtain $\dot{\delta E_i^a}$ from
$\delta E_i^a=-\frac 12\bar{p}h_i^a$, i.e.
\begin{eqnarray}
\dot{\delta E_i^a} &=&-\frac 12\left(
\dot{\bar{p}}h_i^a+\bar{p}\dot{h_i^a}\right)
\nonumber  \label{23} \\
&=&-\frac 12\left( 2\bar{p}\frac{c_h^{(n)}}\gamma \cos \left( \tilde{\mu}%
\gamma \bar{k}\right) \mathfrak{G}_n\left( \tilde{\mu}\gamma
\bar{k}\right) h_i^a+\bar{p}\dot{h_i^a}\right) . \nonumber \\
\end{eqnarray}
From Eqs.(\ref{22}) and (\ref{23}), we have
\begin{equation}
\delta K_a^i=\frac 12\left[ \dot{h_a^i}+\left(
2\frac{c_h^{(n)}}\gamma \cos
\left( \tilde{\mu}\gamma \bar{k}\right) \mathfrak{G}_n\left( \tilde{\mu}%
\gamma \bar{k}\right) -\frac{c_{mh}^{(n)}}\gamma \right)
h_a^i\right] .
\end{equation}
So, the $\dot{\delta K_a^i}$ will be
\begin{eqnarray}
\dot{\delta K_a^i} &=&\frac 12\left[ \ddot{h_a^i}+h_a^i\partial _t\left( 2%
\frac{c_h^{(n)}}\gamma \cos \left( \tilde{\mu}\gamma \bar{k}\right) %
\mathfrak{G}_n\left( \tilde{\mu}\gamma \bar{k}\right) -\frac{c_{mh}^{(n)}}%
\gamma \right) \right.   \nonumber  \label{25} \\
&&\left. +\left( 2\frac{c_h^{(n)}}\gamma \cos \left( \tilde{\mu}\gamma \bar{k%
}\right) \mathfrak{G}_n\left( \tilde{\mu}\gamma \bar{k}\right) -\frac{%
c_{mh}^{(n)}}\gamma \right) \dot{h_a^i}\right] . \nonumber \\
\end{eqnarray}
Again, we can also obtain $\dot{\delta K_a^i}$ from Hamiltonian
equation
\begin{eqnarray}
\dot{\delta K_a^i} &=&\left\{ \delta K_a^i,H_G^Q\right\} +\left\{ \delta
K_a^i,H_{matter}\right\}   \nonumber \\
&=&\left\{ \delta K_a^i,H_{matter}\right\} +\frac 14\left( \frac{c_h^{(n)}}%
\gamma \right) ^2h_a^i-\frac 12\frac{c_{mh}^{(n)}}\gamma \dot{h_a^i}+\frac 12%
\nabla ^2h_a^i  \nonumber \\
&&-\frac 12\frac{c_{mh}^{(n)}}\gamma \left( 2\frac{c_h^{(n)}}\gamma \cos
\left( \tilde{\mu}\gamma \bar{k}\right) \mathfrak{G}_n\left( \tilde{\mu}%
\gamma \bar{k}\right) -\frac{c_{mh}^{(n)}}\gamma \right) h_a^i.\label{26}
\end{eqnarray}
\begin{widetext}
From Eqs.(\ref{25}) and (\ref{26}), one can obtain the gravitational
waves equation
\begin{equation}
\frac 12\left[ \ddot{h_a^i}+2\frac{c_h^{(n)}}\gamma \cos \left( \tilde{\mu}%
\gamma \bar{k}\right) \mathfrak{G}_n\left( \tilde{\mu}\gamma
\bar{k}\right) \dot{h_a^i}-\frac 12\nabla
^2h_a^i+T_Q^{(n)}h_a^i\right] =8\pi G\Pi _{Qa}^i,
\end{equation}
where $\Pi _{Qa}^i$ is the source terms from the matter Hamiltonian
and
\begin{eqnarray}
T_Q^{(n)} &=&-2\frac{\partial \tilde{\mu}}{\partial \bar{p}}\frac{\bar{p}}{%
\tilde{\mu}}\left\{ 2\tilde{\mu}^2\gamma ^2\left[ \left( \frac{c_h^{(n)}}%
\gamma \right) ^4\mathfrak{G}_n\left( \tilde{\mu}\gamma
\bar{k}\right) -\left( \frac{c_h^{(n)}}\gamma \right)
^3\frac{cos^2\left( \tilde{\mu}\gamma \bar{k}\right) }{\gamma
\tilde{\mu}}\mathfrak{B}_n\left( \tilde{\mu}\gamma
\bar{k}\right) \right] \right.   \nonumber  \label{28} \\
&&\left. -\frac{c_h^{(n)}}\gamma \left[ \cos \left( \tilde{\mu}\gamma \bar{k}%
\right) \mathfrak{G}_n\left( \tilde{\mu}\gamma \bar{k}\right) \frac{%
c_{mh}^{(n)}}\gamma -\cos \left( m\tilde{\mu}\gamma \bar{k}\right) %
\mathfrak{G}_n\left( m\tilde{\mu}\gamma \bar{k}\right)
\frac{c_h^{(n)}}\gamma
\right] \right\}   \nonumber \\
&&+\frac 12\left( \frac{c_h^{(n)}}\gamma \right) ^2\left\{ 2\frac{c_h^{(n)}}%
\gamma \gamma \tilde{\mu}\left[ \sin \left( \tilde{\mu}\gamma \bar{k}\right) %
\mathfrak{G}_n\left( \tilde{\mu}\gamma \bar{k}\right) -cos^2\left( \tilde{\mu%
}\gamma \bar{k}\right) \mathfrak{B}_n\left( \tilde{\mu}\gamma
\bar{k}\right)
\right] \right.   \nonumber \\
&&\left. -1+\cos \left( m\tilde{\mu}\gamma \bar{k}\right) \mathfrak{G}%
_n\left( m\tilde{\mu}\gamma \bar{k}\right) \right\} -\left( \frac{c_h^{(n)}}%
\gamma \cos \left( \tilde{\mu}\gamma \bar{k}\right) -\frac{c_{mh}^{(n)}}%
\gamma \right) ^2,
\end{eqnarray}
\end{widetext}
where
\begin{equation}
\mathfrak{B}_n\left( \tilde{\mu}\gamma \bar{k}\right) =\sum_{l=0}^n\frac{%
2l(2l!)}{2^{2l}(l!)^2}\left[ \sin \left( \tilde{\mu}\gamma
\bar{k}\right) \right] ^{2l-1}.  \label{29}
\end{equation}
When $n=0$, $\mathfrak{B}_n\left( \tilde{\mu}\gamma \bar{k}\right)
=0$. So it will not appear in conventional way. The definition of
the effective mass is
\begin{equation}
m_g^2:=\frac{T_Q}{a^2}. \label{mass}
\end{equation}
When $\tilde{\mu}\gamma\bar{k}\rightarrow\frac{\pi}{2}$, $n\rightarrow\infty$%
, the mass term will never vanish. From this, we can confirm that
the nonzero mass of gravitational wave results from the quantum
nature of Riemannian geometry of LQG.

From the definition of gravitational wave \cite{lqct}, we should require $%
T^{(n)}_Q\geq0$, and this is another condition to restrict the range of $%
\beta$.

\section{Lattice refinements}

\label{s5}
In process of obtaining the
range of $\beta$ in \cite{lqct}, the author expand the ``$\sin$" (and ``$\cos$") of Eq.(\ref{19})($n=0$)
and just take the first several terms of it. So in his paper, the first
non-zero term of anomaly part is $k^4$. By this way, one can obtain the
relationship between $m$ and $\beta$ from
Eq.(\ref{19})($n=0$), and obtain the range of $\beta$ by requiring the $%
T^{(n)}_Q>0$ from Eq.(\ref{28})($n=0$).

However, there are some problems in this method. First of all, the relation of $m^2=5+2\beta $ in
\cite{lqct} is obtain by requiring
the term of $k^4$ is vanished. But it can not ensure the whole
anomaly part can be canceled, because there are still $k^6,k^8,etc$. Secondly, when we consider the higher corrections, there will be
some high order terms like $\sin ^3(x)$ appear. If we keep more terms of ``$\sin$" (and ``$\cos$"), we will find
that the first non-zero term is not $k^4$, maybe, it will be $k^5$, it
depend on how many terms you kept. It decides the different relation between $m$ and $%
\beta $. So if we want to obtain the range of
$\beta$ more accurate, we should not expand the ``$\sin$"(and ``$\cos$") in the
equations, in other words, we keep all the terms of it.

From the discussions above we can know that, Eq.(\ref{32}) and $%
T^{(n)}_Q\geq0$ are two restricts to $\beta$, so we analyze these
two restricts respectively. At first, we note that, the product
$\tilde{\mu}\gamma\bar{k}$ always appear
together in the expression of $c_{mh}^{(n)}\tilde{\mu}$ , so we set $x=%
\tilde{\mu}\gamma\bar{k}$, and then $c_{mh}^{(n)}\tilde{\mu}$ is the
function of $x$. When Big Bounce occurs at $x=\frac{\pi}{2}$, and
$x\rightarrow0 $ with the expansion of the universe, we can draw the
graphs of this function between 0 to $x=\frac{\pi}{2}$ with
different value of $n$.

From Fig.\ref{f1} we can see that, when $n=0$ (it correspond the
conventional holonomy corrections), the case of $\beta=-\frac 52$ (which was lower bound in \cite{lqct}) can fill the condition of $c_{mh}^{(n)}%
\tilde{\mu}\leq \frac \pi 2$. But it can be more lower
than that because $\beta =-2.7$ can also fill the condition.
However, we will be concerned about the higher corrections, so let us analyze
the case of large $n$. From Fig.\ref{f1}, we can see that, the
larger $n$ lead to the bigger lower bound of $\beta $.
\begin{widetext}
\begin{center}
\begin{figure}[h!]
\includegraphics[width=0.8\textwidth]{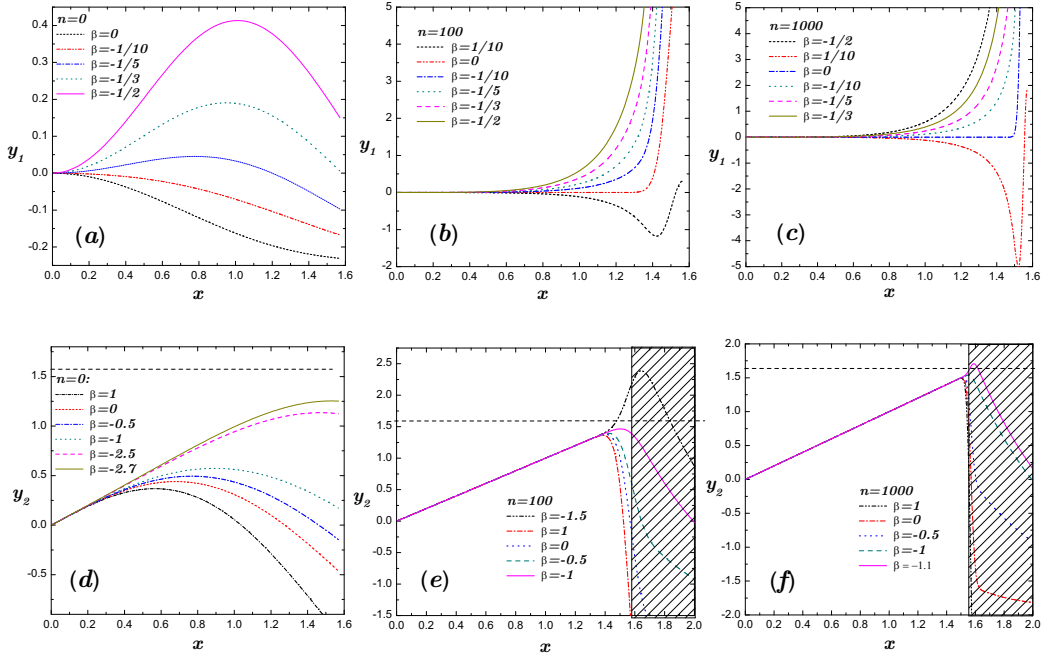}
\caption{The evolution of $y_1=\frac{T^{(n)}_Q}{\bar{k}^2}$ and
$y_2=c_{mh}^{(n)}\tilde{\mu}$ with $x=\tilde{\mu}\gamma\bar{k}$ are
shown in [(a),(b),(c)] and [(d),(e),(f)] respectively. The dash
lines in (d),(e)and (f) are $y_2=\frac{\pi}{2}$.
 }\label{f1}
\end{figure}
\end{center}
\end{widetext}

Because the Eq.(\ref{32}) should be kept everywhere, it includes the
point of the Big Bounce i.e. $x=\frac{\pi}{2}$. Inserting
$x=\frac{\pi}{2}$ to (\ref{32}), it will become
\begin{equation}
\left(\frac{1-2\beta}{2}-\frac{1}{2}\right)\frac{\pi}{2}\leq\frac{\pi}{2},
\end{equation}
which leads to
\begin{equation}  \label{33}
\beta\geq-1.
\end{equation}

On the other hand, the Eq.(\ref{19}) is the relationship of $m$ and
$\beta $. We insert the $\frac{c_{mh}^{(n)}}{\gamma \bar{k}}$ into
Eq.(\ref{28}). On the right hand of Eq.(\ref{28}), there are still
some terms which contain $m$, so we can use the following relation
to replace this terms
\begin{equation}
\cos \left( m\tilde{\mu}\gamma \bar{k}\right) \mathfrak{G}_n\left( m\tilde{%
\mu}\gamma \bar{k}\right) =\tilde{\mu}\gamma \bar{k}\frac \partial
{\partial
\left( \tilde{\mu}\gamma \bar{k}\right) }\frac{c_{mh}^{(n)}}\gamma +\frac{%
c_{mh}^{(n)}}\gamma .  \label{30}
\end{equation}

From Eqs.(\ref{19}), (\ref{28}) and (\ref{30}) we can see that $\frac{%
T_Q^{(n)}}{\bar{k}^2}$ is also the function of $x=\tilde{\mu}\gamma \bar{k}$%
, and there is two parameters $n$ and $\beta $ in it.

From the evolution of $\frac{T^{(n)}_Q}{\bar{k}^2}$ with $n=0$ (see Fig.\ref{f1}(a)),
we can see that, if we require the $\frac{T^{(n)}_Q}{\bar{k}^2}>0$,
the $\beta$ should be smaller than
$-\frac{1}{3}$. With Eq.(\ref{33}), the range of $\beta$ is
$-\frac{1}{3}>\beta\geq-1$. This result is smaller than
$-0.1319>\beta\geq-5/2$. It is because we use the ``$\sin$", not the
first orders of expanding term of ``$\sin$".

When $n>0$, we find that the upper bound of $\beta$ is larger than $-\frac{%
1}{3}$, and when $n\rightarrow\infty$, the upper bound will be zero.
We display the $n=100$ and $n=1000$ in Fig.\ref{f1} also.

So, the final range of $\beta $ should be $\left[ -1,0\right] $.
From this we can see that $\beta =0$ is not eliminated like in
\cite{lqct} and $\beta =-\frac 12$ is also in this range.

\section{Discussion}

\label{s6} In this paper, we apply the higher order holonomy
corrections to
the perturbation theory of cosmology. When we take the limit of $%
n\rightarrow\infty$, the form of the LQC will be back to classical
theory, but the effect of the quantum geometry will be kept. From
the analyses above, we know that the mass of gravitational wave
will not be vanished when $n\rightarrow\infty$. It will decrease to
zero with the expansion of the universe. So it is the ``pure"
quantum effect that the gravitational wave have nonzero mass.

Other important effects are related to the discrete space-time
geometry.
Discrete space means the existence of the area gap, and there is a function $%
\tilde{\mu}(p)$ related to this area gap. The form of function $\tilde{\mu}%
(p)$ has an important impact on LQC. But now the framework of theory
is not perfect to decide this function, so we only can restrict the
form of the function as $\tilde{\mu}\propto p^{\beta}$, from some
other aspects like effective theory and perturbation theory of
cosmology.

In the effective LQC framework, we apply two conditions to limit the
range of $\beta$. One is anomaly free, which means that the
constraint algebra of vector mode should be closed, when we consider
the the quantum effect. It is the mathematical requirements of
the theory. This can restrict $%
\beta $ to be $[ -1,+\infty ) $.

The other condition is the requirements of positive definite mass of
gravitational waves. This is the physical requirement. We can not
ensure that the mass of gravitational wave is positive when $\beta
>0$. And from Fig.\ref{f1}, we can see that the behavior of mass
between $\beta
>0$ and $\beta \leq 0$ is very different. Therefore it can restrict
the range of $\beta $ to be $( -\infty ,0] $. This
requirement seems very natural. However, we do not yet understand
the true meaning of the mass of gravitational waves, so this
condition is only an assumption. The correctness of this assumption
needs to be verified in future studies.

In conclusion, the range of $\beta$ should be $[-1,0]$. But this
range is only decided by perturbation theory of cosmology. It cannot
exclude $\beta=0$. So the excluding of $\beta=0$ is based on the
prediction of theory rather than theory itself. This may not be its
final scope because only two conditions were discussed in this
article. Certainly there are many other conditions to limit the
range of parameter. If we can restrict $\beta$ to a unique value,
say $-1/2$, from the theory itself rather than predictive power of
theory, then the theory will be more self-consistency. So, in the
future studies, we can compare different conditions on the parameter
values to examine the self-consistency of theory.

\acknowledgements The work was supported by the National Natural
Science of China (No. 10875012) and the Fundamental Research Funds
for the Central Universities.

\end{document}